# Porting and optimizing UniFrac for GPUs

Reducing microbiome analysis runtimes by orders of magnitude

Igor Sfiligoi
University of California San Diego
La Jolla CA USA
isfiligoi@sdsc.edu

Daniel McDonald
University of California San Diego
La Jolla CA USA
danielmcdonald@ucsd.edu

Rob Knight
University of California San Diego
La Jolla CA USA
robknight@ucsd.edu

## ABSTRACT

UniFrac is a commonly used metric in microbiome research for comparing microbiome profiles to one another ("beta diversity"). The recently implemented Striped UniFrac added the capability to split the problem into many independent subproblems and exhibits near linear scaling. In this paper we describe steps undertaken in porting and optimizing Striped Unifrac to GPUs. We reduced the run time of computing UniFrac on the published Earth Microbiome Project dataset from 13 hours on an Intel Xeon E5-2680 v4 CPU to 12 minutes on an NVIDIA Tesla V100 GPU, and to about one hour on a laptop with NVIDIA GTX 1050 (with minor loss in precision). Computing UniFrac on a larger dataset containing 113k samples reduced the run time from over one month on the CPU to less than 2 hours on the V100 and 9 hours on an NVIDIA RTX 2080TI GPU (with minor loss in precision). This was achieved by using OpenACC for generating the GPU offload code and by improving the memory access patterns. A BSD-licensed implementation is available, which produces a C shared library linkable by any programming language.

## 1 Introduction

The study of the microbiome has rapidly expanded over the past decade, in part because of the insight afforded by UniFrac [1]. UniFrac is a phylogenetic measure of beta-diversity that assesses differences between pairs of microbiome profiles. UniFrac is useful for microbial community analysis because it can account for the evolutionary relationships between microbes present within a sample. Other distance metrics, such as Euclidean distance, Bray-Curtis, and Jaccard, make the unrealistic implicit assumption that all organisms are equally related, leading to statistical artifacts, particularly with sparse data matrices (these are typical in real-world cases because most kinds of microbes are not found in most locations). Microbiome studies have recently transitioned from experimental designs with a few hundred samples to designs spanning tens of thousands of samples. Large-scale studies, such as the Earth Microbiome Project (EMP) [2], afford the statistics crucial for untangling the many factors that influence microbial community composition.

Having an efficient and scalable implementation for computing UniFrac thus becomes crucial for the advancement of science. A scalable implementation, named Striped UniFrac, was recently proposed and implemented [3]. This algorithm is highly parallelizable and shows almost linear scaling with the number of compute nodes. The existing implementation, however, does not scale linearly with the number of CPU cores on a single node, which is not entirely surprising given its memory-heavy nature.

Massively parallel algorithms, especially memory-heavy ones, are natural candidates for GPU compute. We thus ported the algorithm to GPU resources. One main driver was the desire to avoid CPU-only and GPU-only code paths, to facilitate sustainable long-term support. Using OpenACC [4] was thus a natural choice, because it allows for co-existence of CPU and GPU compute, with conditional creation of GPU offload sections.

Section 2 provides an overview of the minimal changes needed to get the original implementation to execute effectively on GPU resources. Section 3 provides an overview of the additional changes that were put in place to optimize the code to better exploit the GPU architecture. Finally, Section 4 provides an analysis of the obtained results when switching from fp64 to fp32 compute in GPU code, which speeds by 3x the GPU compute on mobile and gaming GPUs.

The original code was compiled with gcc version 7.3, and all OpenACC enabled code was compiled with PGI C compiler version 19.10. All tests were run in Linux environment.

## 2 Porting Striped UniFrac to OpenACC

The most time-consuming part of the original Striped UniFrac implementation is composed of a set of tight loops that operate on adjacent, independent memory cells. Converting such a loop to a GPU offload section with OpenACC is as easy as adding a pragma to the code.

Unfortunately, the original code used a set of memory buffers in such a loop, for perceived efficiency reasons. Passing an array of pointers into an OpenACC sections is, however, not supported in any existing compiler, so some code refactoring was needed. After assessing the usage patterns, we realized that over half of the codebase assumed such a data structure, making proper refactoring a tedious and risky endeavor. As a result, we decided to create a unified temporary buffer for the time-consuming code to operate on, then make a final copy at the end of the computation. While not ideal, the computational cost of this operation is very small, although it does increase the memory footprint of the application. We may clean up the implementation sometime in the future.

With a unified memory buffer in place, it became possible to use simple pointer manipulation math to access the necessary memory cells. This further helped by allowing for fusing of loops and thus increase the available parallelism. Comparison of a subset of the code before and after is provided in Figure 1.

```
for(unsigned int stripe = start;
    stripe < stop; stripe++) {
  dm_stripe = dm_stripes[stripe];
  for(unsigned int j = 0;
      j < n_samples / 4; j++) {
    int k = j * 4;
    double u1 = emb[k];
    double u2 = emb[k+1];
    double v1 = emb[k + stripe + 1];
    double v2 = emb[k + stripe + 2];
    …
    dm_stripe[k]   += (u1-v1)*length;
    dm_stripe[k+1] += (u2-v2)*length;
  }
}
```

```
#pragma acc parallel loop collapse(2) \
        present(emb,dm_stripes_buf)
for(unsigned int stripe = start;
    stripe < stop; stripe++) {
  for(unsigned int j = 0;
      j < n_samples / 4; j++) {
    int idx =(stripe-start_idx)*n_samples;
    double *dm_stripe =dm_stripes_buf+idx;
    int k = j * 4;
    double u1 = emb[k];
    double u2 = emb[k+1];
    double v1 = emb[k + stripe + 1];
    double v2 = emb[k + stripe + 2];
    …
    dm_stripe[k]   += (u1-v1)*length;
    dm_stripe[k+1] += (u2-v2)*length;
  }
}
```

Figure 1: A subset of the most time-consuming code, before and after.

This simple change was all that was needed to compile a working version of UniFrac that could run on a GPU. The GPU runtime of this new executable compared very favorably with the CPU version; computing UniFrac on the EMP sample lasted 1.5 hours on an NVIDIA Tesla V100 GPU, versus 13 hours using an Intel Xeon E5-2680 v4 CPU, using all 14 cores concurrently.

## 3 Optimizing for GPU compute

Having achieved a working GPU port, we analyzed its performance. The first issue was the partial manual unrolling of the inner loop in the above code; that was done to help the CPU compiler generate better vector instructions. Unfortunately, it resulted in a striped memory access pattern in the GPU code, because the compiler automatically generated vector code based on the loop itself. Removing the manual unrolling, the time needed for the same compute was reduced from 92 minutes to 64 minutes on the NVIDIA Tesla V100 GPU.

Further code analysis pointed to the likely bottleneck to be the repeated updating of the main memory buffer. The original logic would retrieve, in the CPU section, one input buffer per GPU kernel invocation, which would then extract and transform the needed information and finally updated the main memory buffer. The same operation would be performed on O(10k) input buffers in sequence. This is suboptimal for two reasons: first, writing to memory is significantly more expensive than reading from it, and second, each kernel invocation has a non-negligible overhead.

The implemented solution was to batch many input buffers in a single kernel invocation and modify the loops to process the data from all the input buffers before updating the main memory buffer. This slightly increased the memory footprint of the application, but resulted in a further reduction in runtime to about 33 minutes on the NVIDIA Tesla V100 GPU. The updated code snippet is available in Figure 2.

```
#pragma acc parallel loop collapse(2) \
        present(emb,dm_stripes_buf,length)
for(unsigned int stripe = start;
    stripe < stop; stripe++) {
  for(unsigned int k = 0;
      k < n_samples ; k++) {
    double my_stripe = dm_stripe[k];
#pragma acc loop seq
    for (unsigned int e=0;
         e<filled_embs; e++) {
      uint64_t offset = n_samples*e;
      double u = emb[offset+k];
      double v = emb[offset+k+stripe+ 1];
      my_stripe += (u-v)*length[e];
    }
    …
    dm_stripe[k]   += (u1-v1)*length;
  }
}
```

Figure 2: A subset of the most time-consuming code after input buffer batching.

Once the above changes were in place, it became obvious that the same input buffers were accessed multiple times during the execution of a single GPU kernel. The access pattern was however such that the next reuse came only at a much later time, trashing the cache. We thus proceeded in splitting the main loop in such a way that it maximized both vectorization opportunities and input buffer cache locality, as can be seen in Figure 3. Note that it is very important to properly align the memory buffers and pick the right value for the grouping parameters, as it can drastically affect the observed run time.

```
#pragma acc parallel loop collapse(3) \
        present(emb,dm_stripes_buf,length)
for(unsigned int sk = 0;
    sk < sample_steps ; sk++) {
  for(unsigned int stripe = start;
      stripe < stop; stripe++) {
    for(unsigned int ik = 0;
        ik < step_size ; ik++) {
      unsigned int k = sk*step_size + ik;
      …
      double my_stripe = dm_stripe[k];
#pragma acc loop seq
      for (unsigned int e=0;
          e<filled_embs; e++) {
        uint64_t offset = n_samples*e;
        double u = emb[offset+k];
        double v = emb[offset+k+stripe+ 1];
        my_stripe += (u-v)*length[e];
      }
      …
      dm_stripe[k]   += (u1-v1)*length;
    }
  }
}
```

**Figure 3: A subset of the most time-consuming code in its final incarnation.**

With the latest change, computing UniFrac on the EMP sample took only 12 minutes on the NVIDIA Tesla V100 GPU. We also compiled the new code with OpenACC disabled, computed UniFrac on the EMP sample on the Intel Xeon E5-2680 v4 CPU, using all 14 cores concurrently, and it finished in 193 minutes.

To summarize, Table 1 provides the times needed to compute UniFrac on the EMP sample, both using the original code and after being ported to the GPU. As can be seen, an NVIDIA Tesla V100 GPU provides an order of magnitude improvement over the tested Intel Xeon E5-2680 v4 CPU. Note that the quoted CPU time is for the whole chip, i.e. using all its resources, not single-threaded.

*Table 1: Runtimes of Striped UniFrac on EMP dataset. In chip minutes.*

| Intel Xeon E5-2680 v4 CPU | | NIVIDA Tesla V100 GPU | |
|---|---|---|---|
| Original | Final | OpenACC base | Final |
| 800 | 193 | 92 | 12 |

To verify that the obtained improvements in run time were not specific to the chosen input dataset, we also computed UniFrac on the same input dataset used in [3], which is much bigger and contains 113,721 samples. This input dataset is too big to be ran on a single CPU in reasonable time, so we distributed the compute over several CPUs and GPUs. Using 128 chips in parallel gave us a reasonable per-chip runtime for the CPU systems. Note that running larger subproblems on the GPUs results in a significant speedup, so we ran the GPU compute also with 4 parallel chunks. As can be seen from Table 2, which provides the times needed to compute UniFrac on those 113,721 samples, the GPU version provides several orders of magnitude speedup, both in terms of per-chip and total compute time. It would now be possible to compute UniFrac on 113,721 samples in a couple of hours using a single NVIDIA Tesla V100 GPU.

*Table 2: Runtimes of Striped UniFrac on 113,721 samples. In chip hours.*

|  | Org. CPU version | Final GPU version | |
|---|---|---|---|
|  | 128x E5-2680 v4 | 128x V100 | 4x V100 |
| **Per chip** | 6.9 hours | 0.23 hours | 0.34 hour |
| **Aggregated** | 890 hours | 30 hours | 1.9 hours |

## 4 Validating 32-bit floating point compute

UniFrac was originally designed and always implemented using fp64 math operations. The higher-precision floating point math was used to maximize reliability of the results.

On CPU cores, the penalty to pay for fp64 versus fp32 is expected to be small. Only a fraction of the total compute uses floating point compute, and that part would be at best 2x faster on CPUs. The situation is similar on server-class GPUs, like the NVIDIA Tesla V100 GPU, but on mobile and gaming GPUs fp64 compute is 32x slower than fp32 compute. We measured the time needed to compute UniFrac, on both the EMP dataset and the dataset containing 113,721 samples, using the latest GPU-enabled UniFrac code, and we observe between 2x and 6x speedup in fp32-bit mode, see Table 3 and Table 4. Note that the compute times on the CPU were virtually identical for the fp32 and fp64 code paths.

*Table 3: Runtimes of the final GPU-enabled Striped Unifrac on EMP, using fp64 vs fp32 math. All GPUs by NVIDIA. In minutes.*

|  | V100 | 2080TI | 1080TI | 1080 | Mobile 1050 |
|---|---|---|---|---|---|
| **fp64** | 12 | 59 | 77 | 99 | 213 |
| **fp32** | 9.5 | 19 | 31 | 36 | 64 |

*Table 4: Runtimes of the final GPU-enabled Striped Unifrac on 113,721 samples, using fp64 vs fp32 math. Using multiple GPUs by NVIDIA. In aggregated hours.*

|  | V100 | 2080TI | 1080TI |
|---|---|---|---|
| **fp64** | 1.9 hours | 49 hours | 67 hours |
| **fp32** | 1.3 hours | 8.5 hours | 22 hours |

Getting unusable results fast would, however, not be helpful. We thus compared the results of the compute using fp32-enabled and fp64-only code, using the same EMP input, and observed a near identical result (Mantel $R^2$ 0.99999; p<0.001, comparing pairwise distances in the two matrices). The fp32-enabled code can thus be used for most microbiome discovery work, especially if run on personal equipment, with the fp64-only code used only in the unusual situation where the relative abundances of the input data or the tree branch lengths exhibit a very high dynamic range in elements of the distance matrix that contribute substantially to downstream results, e.g. after dimensionality reduction.

## 6  Conclusions

Microbiome studies are transitioning from experimental designs with a few hundred samples to much larger designs spanning tens of thousands of samples. Having access to effective but also fast compute tools is thus becoming essential. UniFrac has long been an important tool in microbiome research, and our work now allows many analyses which were previously relegated to large compute clusters to be performed with much lower resource requirements. Even the largest datasets currently envisaged could be processed in reasonable time with just a handful of server-class GPUs, while smaller but still sizable datasets like the EMP can be processed even on GPU-enabled workstations.

We used OpenACC to port the existing Striped UniFrac implementation to GPUs, because this allows a single codebase for both CPU and GPU code, thus significantly reducing long term support burden. Some refactoring of the code was needed to obtain maximum performance from the GPUs, but this refactoring was mostly limited to the most time-consuming part of the code. The increased memory footprint is slightly increased, but we believe this trade-off is well worth the order of magnitude speed improvement in run times.

Finally, we explored the use of lower-precision floating point math to effectively exploit consumer-grade GPUs, which are typical in desktop and laptop setups. We conclude that fp32 math yields nearly identical results to fp64, and should be adequate for the vast majority of studies, making compute on GPU-enabled personal devices, even laptops, a sufficient resource for this otherwise rate-limiting step for many researchers.

## ACKNOWLEDGMENTS

This work was partially funded by US National Science Foundation (NSF) under grants OAC-1826967, OAC-1541349 and CNS-1730158, and by US National Institutes of Health (NIH) under grant DP1-AT010885.